\documentclass[a4paper,11pt]{article}
\usepackage{amsmath, amssymb, amsfonts,indentfirst,graphicx,color,anysize,cite,caption}
\usepackage[abs]{overpic}
\marginsize{3cm}{3cm}{2cm}{2cm}
\title{
	\includegraphics[width=0.35\textwidth]{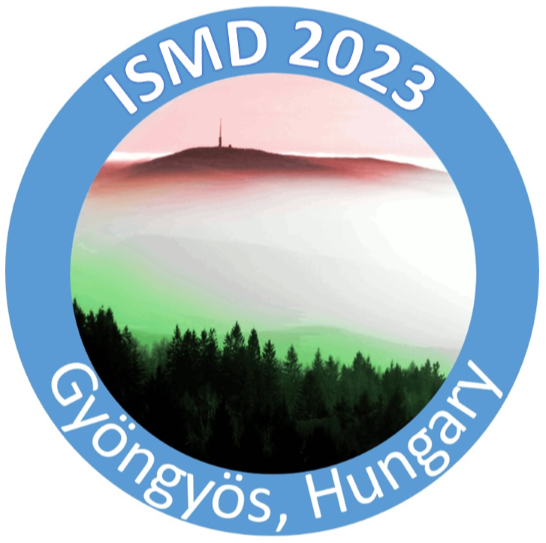}\\[1cm]
	\textbf{Strangeness production in double gap events in ALICE}}
\author{{R. Schicker$^{1}$}\\ (on behalf of the ALICE collaboration)\\ [1ex]
	$^1$Physikalisches Institut, Im Neuenheimer Feld 226, Heidelberg, Germany\\
}
\begin{document}

\maketitle

\begin{abstract} 
  The ALICE detector at the LHC was upgraded in the long shutdown
  of 2019-2021 in order to take data at much-increased Run 3 and 4 rates. 
  The various challenges of this upgrade are presented, and the
  first results of strangeness production in double gap events collected in
  2022 are shown by  presenting distributions of kaon pairs.
\end{abstract}

\section{Introduction}

In the long shutdown of 2019-2021, the ALICE experiment was upgraded in order
to take data at much-increased rates in Runs 3 and 4 \cite{ALICEUP}.
The readout was changed to continuous mode, with the data collected in
proton-proton collisions at a typically 1 MHz rate and subsequently
being processed to identify the events of interest. The selected events are
permanently stored for future processing. All data taken in heavy-ion
collisions at up to 50 kHz are permanently stored, and available for
later analysis.

\section{The ALICE detector upgrades}

The Time-Projection-Chamber (TPC) is not only the main tracking detector of
the ALICE experiment  but also provides information on the specific ionisation 
loss dE/dx for particle identification. The ionisation electrons drift in the
electric field of 400 V/cm to the endplates of the cylindrical field cage.
The  ionisation charge was amplified in Runs 1,2 by multiwire proportional
chambers (MWPCs), with the backflow of positive ions being controlled by
gating grids. The effective rate was limited in Runs 1,2 to $\sim$ 1 kHz.
In order to increase the rate capability to $\sim$ 50 kHz, the MWPCs were
replaced by a stack of \mbox{4 Gas} Electron Multipliers (GEMs) with continuous
readout and synchronous data processing \cite{TPCup}.

The readout rate of the Inner Tracking System (ITS) in Runs 1,2 was limited
\mbox{to 1 kHz.} The Run 3 newly designed system features improved
resolution, less material and faster readout. The new system consists of
7 layers of Monolithic Active Pixel Sensors, with a spatial resolution
(r$\phi$ x z) improved from 11 x 100 ($\mu$m$^{2}$) of the old ITS
to  5 x 5 ($\mu$m$^{2}$). The readout rate of this new ITS system is
100 kHz in Pb-Pb, and 200 kHz in pp collisions \cite{ITSup}.

Outside of the central barrel pseudorapidity range $|\eta| < $ 0.9, a Fast
Interaction Trigger (FIT) system was built for Run 3 to provide precise
collision time information for time-of-flight-based particle information, for
online luminosity monitoring and measuring forward multiplicity \cite{FITup}.
This FIT system consists of the following detectors:

\begin{itemize}
\item {\bf FT0:} Two Cherenkov arrays for minimum bias triggering,
  for determining the collison time and for vertex position calculation.
\item {\bf FV0:} A large scintillator array consisting of five rings and eight
  sectors is positioned on the opposite side of the ALICE muon spectrometer.
  In conjunction with FT0, it is used for centrality and event plane
  determination in heavy-ion collisions.
\item {\bf FDD} (Forward Diffractive Detector): This double-sided scintillator
  array is essential for tagging diffractive events by establishing
  rapidity gaps in the event.
\end{itemize}
\begin{table}[h]
\begin{center}
  \begin{tabular}{p{0.18\textwidth}|p{0.12\textwidth}|p{0.12\textwidth}|p{0.10\textwidth}}
  Detector & $\eta_{min}$ & $\eta_{max}$ & z [cm]\\
  \hline
  FDD-A & 4.8 & 6.3 & 1696.0\\
  FT0-A & 3.5 & 4.9 & 334.6\\
  FV0   & 2.2 & 5.1 & 320.8 \\
  FT0-C & -3.3 & -2.1 & -84.3 \\
  FDD-C & -7.0 & -4.9 & -1956.6\\
\end{tabular}  
\caption{The pseudorapdity range and the z-position of the FIT subsystems.}
\label{table1}
\end{center}
\end{table}
\vspace{-0.2cm}
The parameters of the different subsystems of the FIT detector are shown
in Table \ref{table1}.

\section{The computing system upgrade}

After the upgrade of the ALICE detector systems in the long shutdown of
2019-2021, the raw data flow to the data acquisition system  increased a
hundredfold, up to 3.5 TB/s. A new Online/Offline Computing system, O$^{2}$,
was developed to cope with this challenge \cite{O2up}. The O$^{2}$ system
incorporates continuous readout of most subdetectors, data compression using
partial synchronous  reconstruction and calibration, and the sharing of common
computing resources during and, for asynchronous reconstruction, after data
taking. The reconstructed data are written to disk while the raw data are
discarded. The O$^{2}$ architecture consists
of two major computing layers, the First Level Processors (FLPs) and
the Event Processing Nodes (EPNs). Both layers are highly heterogeneous,
with specialized acquisition cards embedding FPGAs on the FLPs, and
GPUs on the EPNs. The raw data rate of 3.5 TB/s is reduced by the
FLP layer to $\sim$ 635 GB/s  and by the EPN layer to \mbox{100 GB/s.}
This compressed data stream is stored and, after later asynchronous
processing, distributed to the Tier 0 and Tier 1 analysis facilities.

This upgrade of the ALICE experiment resulted in a considerably increased
data taking capability. The double gap data sample from Run 2
amounts to about 10 pb$^{-1}$. In \mbox{Run 3,} a central  barrel data sample of
$\sim$ 10 pb$^{-1}$ is recorded in 6 weeks.

\section{Central diffractive production at the LHC}

Central diffractive events are characterized by particle production at
midrapidity and by the absence at forward and backward rapidities. The ALICE
experiment is ideally suited for measuring such events due to the excellent
global tracking in the central barrel by the combined information of TPC and
ITS and the superb particle identification capability of the TPC. The
absence of particle tracks at forward/backward rapidities can be established
by requiring no signals in the FIT detector systems.

\begin{figure}[h]
\begin{overpic}[width=.32\textwidth]{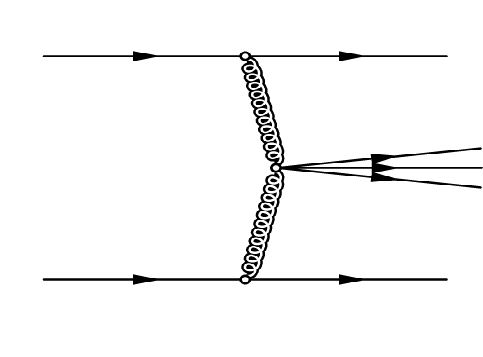}
  \put(30.,-.6){\footnotesize central prod.}
\end{overpic}
\hspace{.1cm}
\begin{overpic}[width=.32\textwidth]{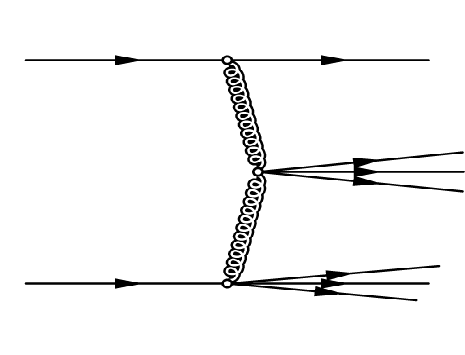}
\put(20.,-.6){\footnotesize central prod./single diss.}
\end{overpic}
\hspace{0.1cm}
\begin{overpic}[width=.32\textwidth]{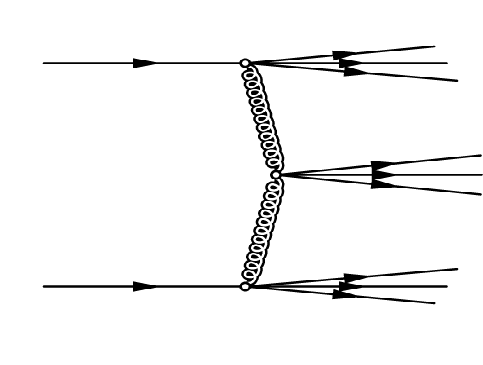}
\put(20.,-.6){\footnotesize central prod./double diss.} 
\end{overpic}
\caption{The event topologies of central production (left), central
  production with single dissociation (middle), and central
production with double dissociation (right).}
\label{fig1}
\end{figure} 

In Figure \ref{fig1}, the topologies of central production events are shown.
In such events, one or both protons can undergo a diffractive excitation, and
subsequently break up into very forward scattered fragments. Central production
at the energies of the LHC is dominated by the Pomeron-Pomeron exchange.
In QCD, the Pomeron is thought to be a dominantly gluonic object.
The study  of particle yields in this gluonic environment, and the comparison
to the corresponding yields in minimum bias collisions is of high
interest in the search for QCD exotica such as hybrids and glueballs.

      \begin{figure}[h]
    \begin{overpic}[width=.54\textwidth]{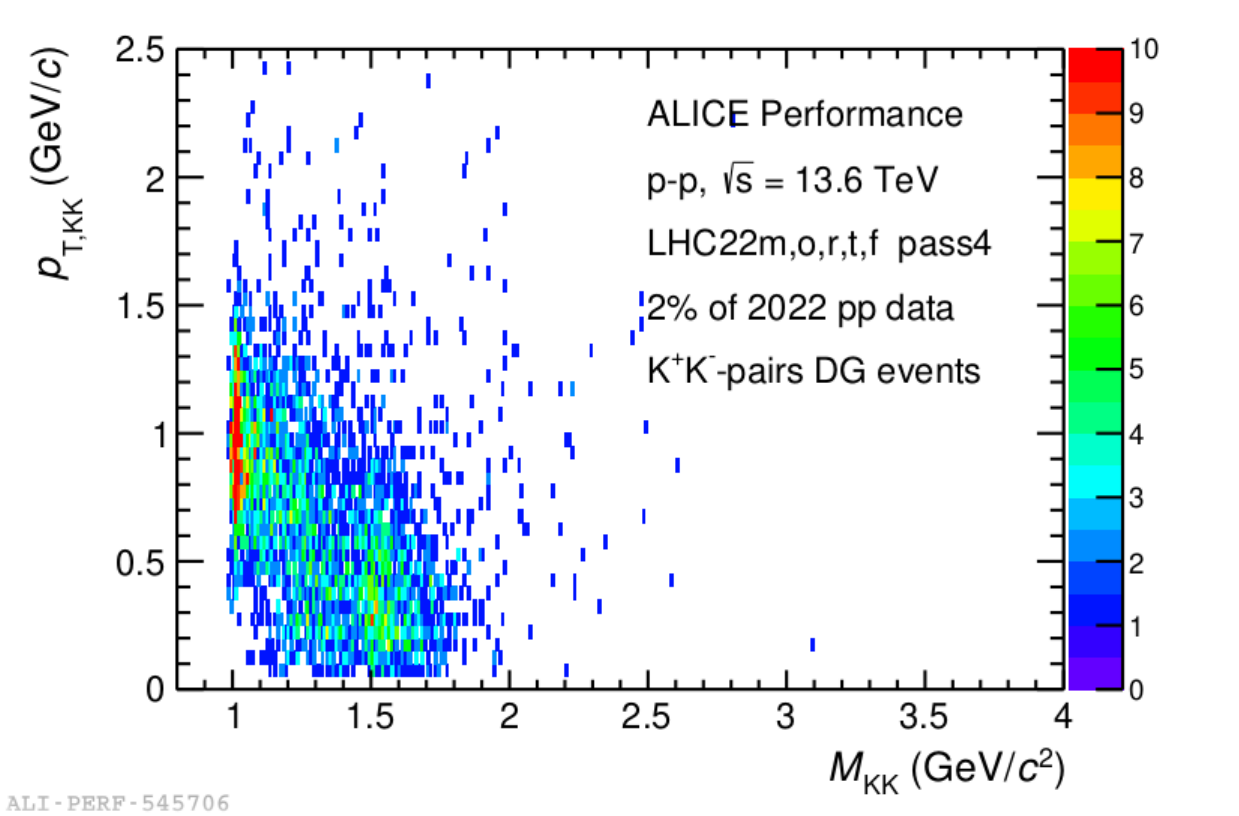}
    \end{overpic}
    \hspace{-.3cm}
    \begin{overpic}[width=.54\textwidth]{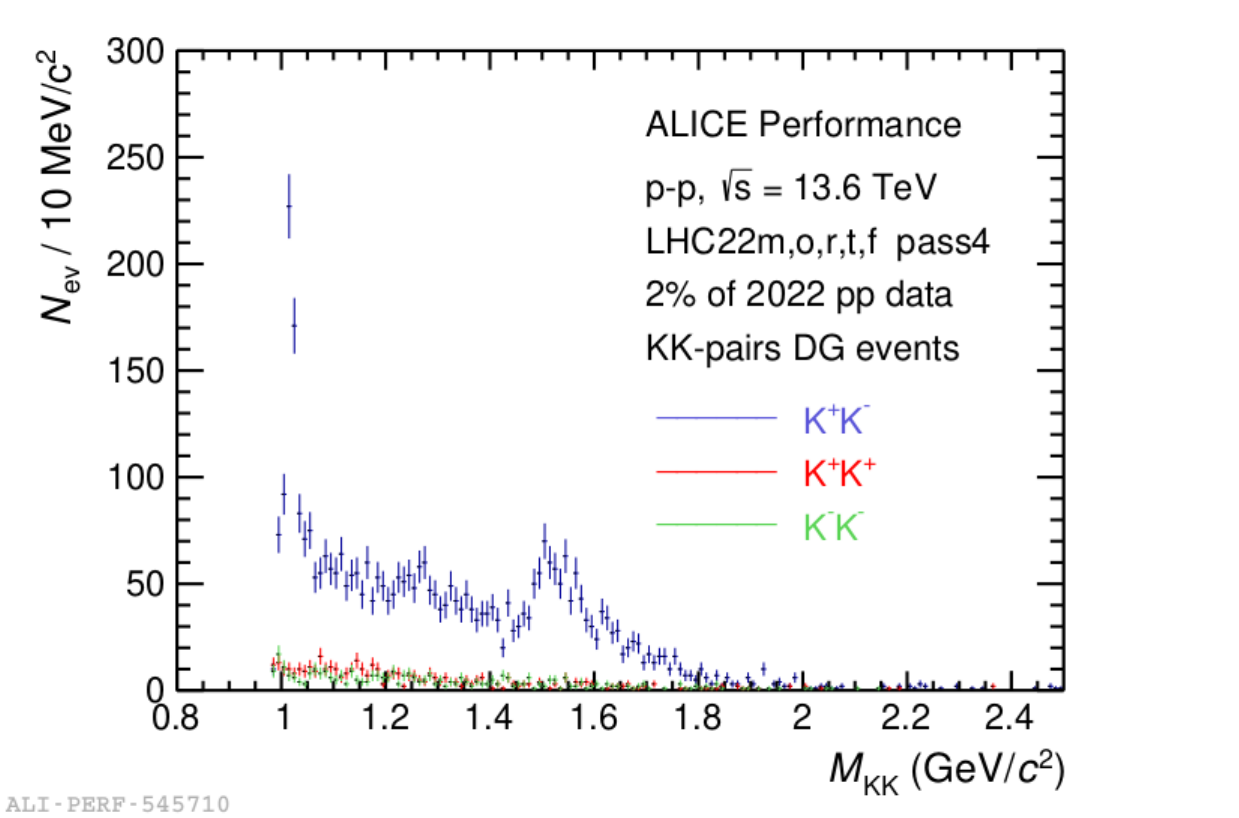}
    \end{overpic}
    \caption{The mass of unlike-sign kaon pairs vs. pair tranverse momentum
      (left), the mass distribution for like and unlike-sign kaon pairs
      (right).}
    \label{fig2}
      \end{figure}

      In Fig. \ref{fig2} on the left, the invariant mass of $K^{+}K^{-}$
      pairs in double gap events is shown as a function of pair transverse
      momentum. The mass distribution shown on the right is derived by
      integrating the 2-D distribution over transverse momentum. The mass
      distribution is shown for unlike and like-sign pairs, indicating
      contamination below the 10\% level for the $K^{+}K^{-}$ signal events.
      The data sample shown in Fig. \ref{fig2} represents about 2\% of the
      statistics taken in 2022.  

      Clearly visible in Fig. \ref{fig2} on the right are peak structures just
      above the two kaon threshold, and at a mass M$_{KK}$ $\approx$ 1.52
      GeV/c$^{2}$. These two structures are tentatively identified as the
      $\phi$(1020) and the f$^{'}_{2}$(1525), that has a known kaon branching
      \mbox{ratio of 88\%.} The limited statistics analysed so far preclude the
      identification of further peaks.
      The ALICE data sample taken in  2022-2023 is larger by a
      factor of about 80 as compared to the sample shown in Fig. \ref{fig2}.
      The analysis of the full-data sample will hence be able to
      give much-improved information on the existence of further structures
      in the $K^{+}K^{-}$ system. 
   
      \section{A model of $q\bar{q}$ bound states}

      Nonrelativistic quark potential models have been applied successfully to
      heavy quark systems. Such models rely on a Coulomb-plus-linear potential
      expected from QCD. The extension to the strangeness and
      light quark (u,d) sectors, however, necessitates the inclusion of
      relativistic effects. A unified framework for light and heavy
      mesons has been presented in Ref. \cite{GI}. This framework relies on a
      relativistic potential which includes a confining part, a spin-orbit
      interaction term, a hyperfine interaction term and an annihilation term.
      The meson spectrum is derived by calculating the bound states in this
      relativistic potential, and is given in spectroscopic notation
      $n^{2S+1}L_{J}$, with $n$ the radial quantum number, $S$ the spin, $L$ the
      orbital angular momentum and $J$ the total angular momentum.

\begin{table}[h]
\begin{tabular}{p{0.10\textwidth}|p{0.15\textwidth}|p{0.12\textwidth}|p{0.10\textwidth}|p{0.15\textwidth}|p{0.15\textwidth}}
\small{$n\:^{2S+1}L_{J}$}  & \small{mass} & \small{PDG} & $J^{PC}$ & \small{mass} (PDG)& \small{width} (PDG) \\
\hline
\small{$1^{3}S_{1}$}&\small{1020} & \small{$\phi$}& 1$^{--}$ & \small{1019} & \small{4}  \\
\small{$1^{3}P_{2}$} &\small{1530} & \small{$f_{2}^{'}$ (1525)}& 2$^{++}$ & \small{1518} & \small{86}  \\
\small{$1^{3}D_{3}$}&\small{1900} & \small{$\phi_{3}$}& 3$^{--}$  & \small{1854} & \small{87}  \\
\small{$1^{3}F_{4}$}&\small{2200} & \small{??} &\small{??} & \small{??} & \small{??}  \\
\end{tabular}   
\caption{Mass and width (in MeV) of isoscalar states with hidden strangeness.}
\label{table2}
      \end{table}

In Table \ref{table2}, the orbital excitations of isoscalar states with hidden
strangeness are listed. The 1st and 2nd columns characterize
the bound states in spectroscopic notation and give the bound state mass
as calculated in Ref. \cite{GI}.
The 3rd and 4th columns list the associated Particle Data Group (PDG)
resonance name, and specify the $J^{PC}$ quantum numbers.
Columns 5 and 6 list the PDG resonance mass and width, respectively.
The PDG does not list any state which could be associated with the F-wave
state as predicted by Ref. \cite{GI}.

\section{Non-linear complex Regge trajectory}

The small but existing non-linear dependence of the total angular momentum
of a resonance as a function of its mass squared can be used to define a Regge
\mbox{trajectory $\alpha(M^{2})$,} a complex entity with real and imaginary
parts being related by a dispersion \mbox{relation \cite{Disp}.}
In a simple model, the imaginary part is chosen as a sum of single
threshold terms,

 \begin{equation}
\Im m\: \alpha(s)\!=\!\sum_{n}\!c_{n} (s\!-\!s_{n})^{1/2} 
\big(\frac{s\!-\!s_{n}}{s}\big)^{\!|\Re e\:\alpha(s_{n})|} \theta(s\!-\!s_{n}).
\label{eq:imag}
 \end{equation}

  In Eq. \ref{eq:imag}, the coefficients $c_{n}$ are fit parameters,
  the parameters $s_{n}$ represent kinematical thresholds of decay
  channels, and $s$ denotes the centre-of-mass energy\cite{RSdiff22}. 

  The real part is defined by the value of the total angular momentum,
  whereas the imaginary part is related to the decay width $\Gamma$,

\begin{equation}
 \Im m\: \alpha(M^{2}_{R}) = \Gamma(M_{R})\:\alpha^{'}(M_{R})\:M_{R}. 
\label{eq:width}
\end{equation}

In Eq. \ref{eq:width}, the relation between the imaginary part
$\Im m\: \alpha(M^{2})$ and the width function $\Gamma(M)$ of the trajectory
is shown, with  $\alpha^{'}(M)$ denoting the derivative of the real part.

  \subsection{The ($\phi,f^{'}_{2}$)-trajectory}

  A fit to the isoscalar states $\phi,f^{'}_{2}, \phi_{3}$ shown in Table \ref{table2} defines the
  ($\phi,f^{'}_{2}$)-trajectory. For the fit, the PDG values for the masses and widths of $\phi,f^{'}_{2}, \phi_{3}$
  resonances are used.

  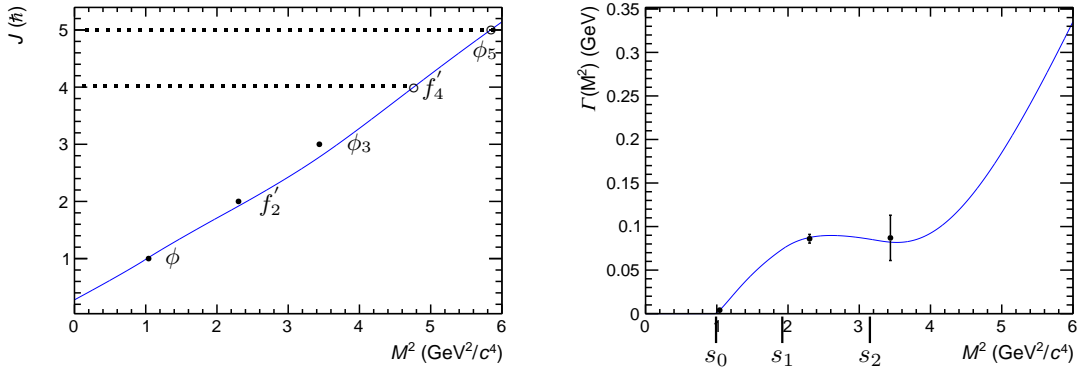
\begin{figure}[h]
    \begin{overpic}[width=.50\textwidth]{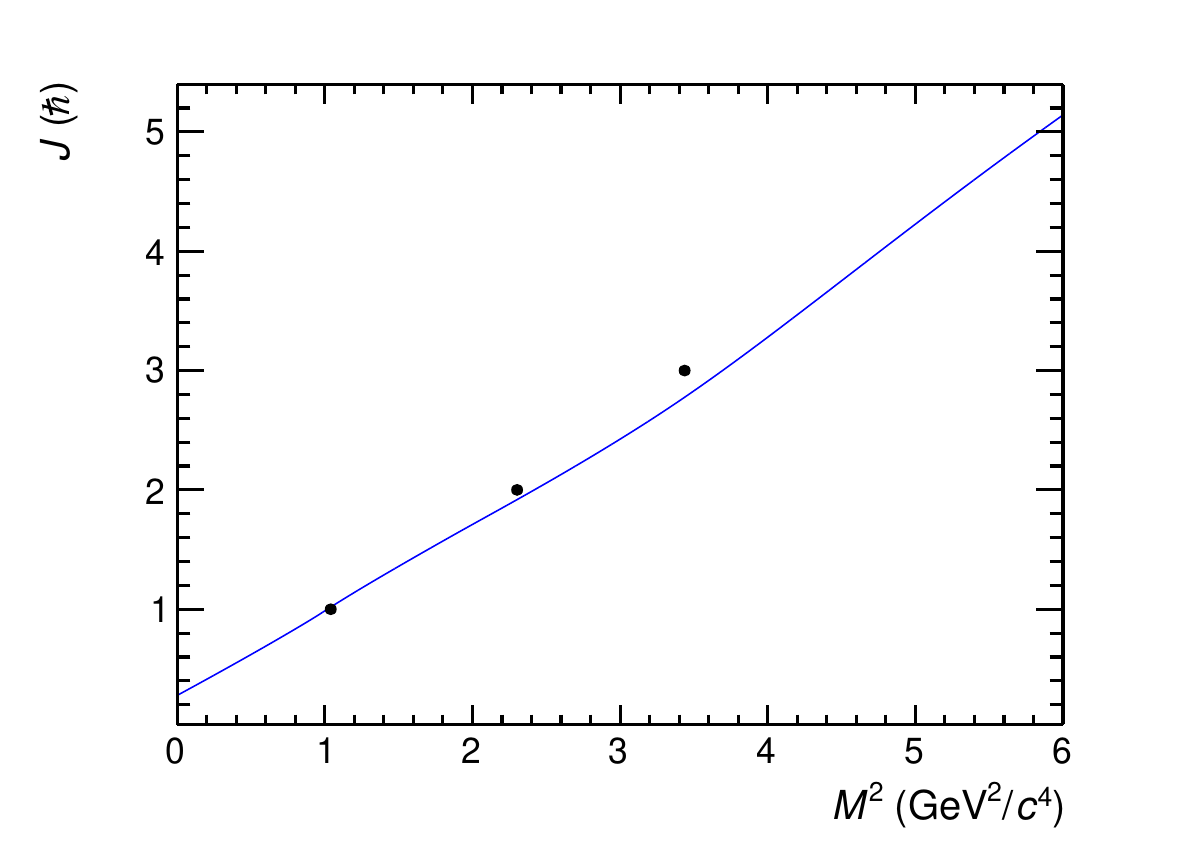}

\multiput(35.,108.8)(4.0,0){31}{\color{black}\linethickness{0.4mm}\line(1,0){1.2}}
\multiput(36.0,130.)(4.0,0){39}{\color{black}\linethickness{0.4mm}\line(1,0){1.2}}

\put(159.,108.4){\color{black}\circle{3.0}}
\put(188.,130.){\color{black}\circle{3.0}}

\put(162.,105.6){\color{black}{\footnotesize{$f_{4}^{'}$}}}
\put(181.,120.0){\color{black}{\footnotesize{$\phi_{5}$}}}
\put(66.,42.){\color{black}{\footnotesize{$\phi$}}}
\put(101.,62.){\color{black}{\footnotesize{$f_{2}^{'}$}}}
\put(134.0,85.){\color{black}{\footnotesize{$\phi_{3}$}}}
    \end{overpic}
\hspace{-.2cm} 
\begin{overpic}[width=.50\textwidth]{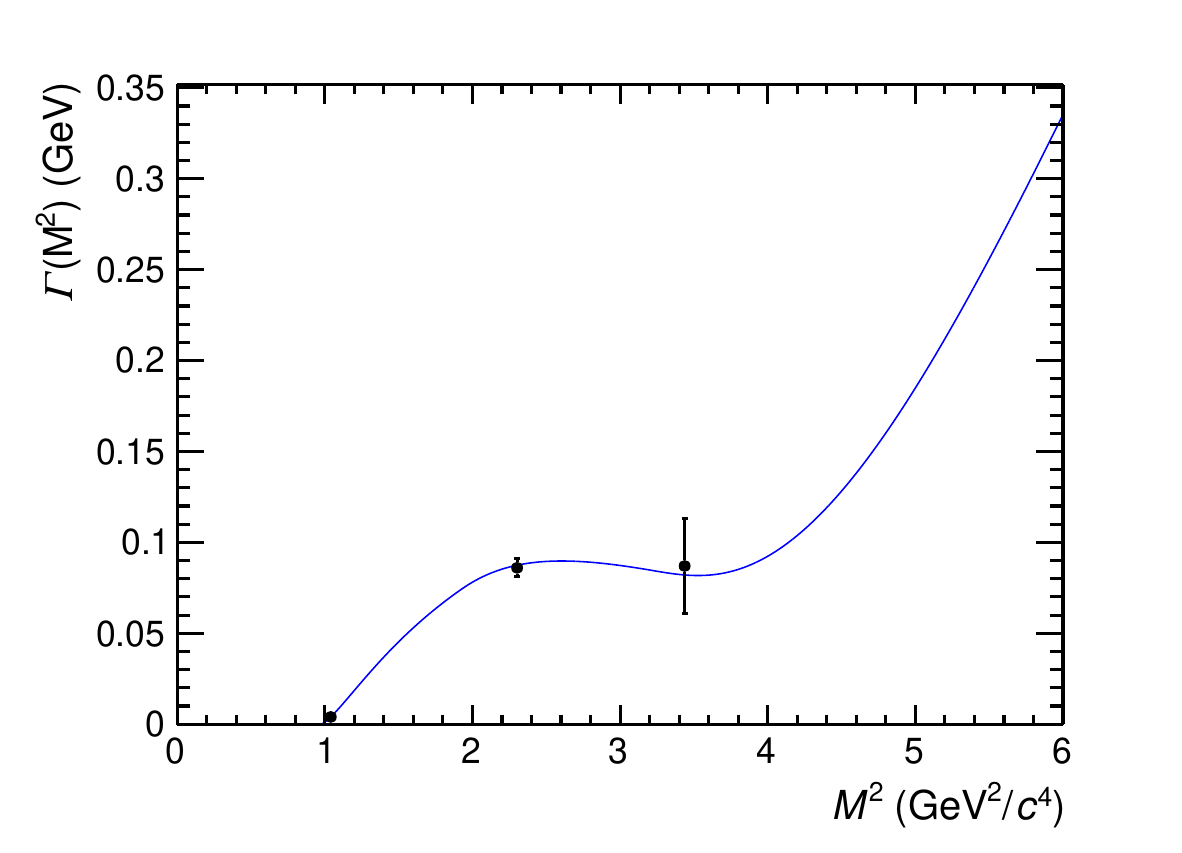}

  \put(58.4,12.){\color{black}\linethickness{0.3mm}\line(0,1){10.0}}
  \put(83.2,12.){\color{black}\linethickness{0.3mm}\line(0,1){10.0}}
  \put(116.,12.){\color{black}\linethickness{0.3mm}\line(0,1){10.0}}
  
\put(54.0,5.){\small \color{black}$s_{0}$}
\put(78.8,5.){\small \color{black}$s_{1}$}
\put(111.6,5.){\small \color{black}$s_{2}$}
\end{overpic}
\caption{The fit of the ($\phi,f^{'}_{2}$)-trajectory.}
\label{fig3}
  \end{figure}

  In Fig. \ref{fig3}, the fit to the real part of the
  ($\phi,f^{'}_{2}$)-trajectory is shown on the left, and the fit to the
  width function $\Gamma(M^{2})$ on the right. The thresholds used in this
  fit are defined by the decays $\phi \rightarrow K\bar{K}$
  with $s_{0}$ = 0.97 GeV$^{2}$, $f^{'}_{2} (1525) \rightarrow KK^{*}$ with
  $s_{1}$ = 1.92 GeV$^{2}$ and $\phi_{3} \rightarrow K^{*}\bar{K}^{*}$
  with $s_{2}$ = 3.18 GeV$^{2}$. This fit predicts the existence of a
  $f^{'}_{4}$ state with mass of 2182 MeV/c$^{2}$ and width of 156 MeV,
  and of a $\phi_{5}$ state with mass of 2417 MeV/c$^{2}$ and
  width of 310 MeV.

  \section{Conclusions and outlook}

  The upgrade of the ALICE detector systems during the shutdown 2019-2021
  has resulted in a hundredfold increase of data rate capability.
  A first analysis of strangeness in double gap events in proton-proton
  collisions shows clear evidence for the strangeonia states $\phi$(1020)
  and $f^{'}_{2}$(1525). The data sample taken by ALICE in 2022-2023 is about
  a factor of 80 larger than the data sample used in the present analysis.
  This much larger sample will allow the search for two not yet
  experimentally identified strangeonia states, the $f^{'}_{4}$(2182) and
  the $\phi_{5}$(2417). The analysis presented here can be extended to the
  study of $\pi K$ pairs, which allows access to the $(u,d)\bar{s}$
  kaonium and the $(\bar{u},\bar{d})s$ antikaonium sector.
  
  \section{ACKNOWLEDGMENTS}
This work is supported by the German Federal Ministry of Education and 
Research under promotional reference 05P21VHCA1.

\end{document}